# Parametrization of the Fe-O$_{water}$ cross-interaction for a more accurate Fe$_3$O$_4$/water interface model and its application to a spherical Fe$_3$O$_4$ nanoparticle of realistic size

Paulo Siani[1], Enrico Bianchetti[1], Hongsheng Liu[1,2] and Cristiana Di Valentin[1,*]

1. Dipartimento di Scienza dei Materiali, Università di Milano-Bicocca, via R. Cozzi 55, 20125 Milano Italy.
2. Laboratory of Materials Modification by Laser, Ion and Electron Beams, Dalian University of Technology, Ministry of Education, Dalian 116024, China

**Abstract**

The accurate description of iron oxides/water interfaces requires reliable force field parameters that can be developed through the comparison with sophisticated quantum mechanical calculations. Here a set of CLASS2 force field parameters is optimized to describe the Fe-O$_{water}$ cross interaction through comparison with hybrid density functional theory (HSE06) calculations of the potential energy function for a single water molecule adsorbed on the Fe$_3$O$_4$ (001) surface and with density functional tight binding (DFTB+U) molecular dynamics simulations for a water tri-layer on the same surface. The performance of the new parameters is assessed through the analysis of the number density profile of a water bulk (12 nm) sandwiched between two magnetite slabs of large surface area. Their transferability is tested for the water adsorption on the curved surface of a spherical Fe$_3$O$_4$ nanoparticle of realistic size (2.5 nm).

---

[*] E-mail address: cristiana.divalentin@unimib.it

## 1. Introduction

Iron oxides/water interfaces are involved in many fundamental and technological processes, therefore accurate force field parameters for the description of the bond between surface iron sites and water oxygens, which we provide through this work, are critical to perform useful molecular dynamics simulations in this fast-developing research field.

Water adsorption on the low-index (001) $Fe_3O_4$ facet has been extensively investigated in the past both through experiments and theoretical calculations.[1,2,3,4,5,6,7,8,9,10,11] In particular, a mixed undissociated/dissociated adsorption mode was determined for one water monolayer adsorbed on the surface by means of a combined experimental and computational study.[11] On half of the metal adsorption sites, water molecules dissociate forming a partially hydroxylated surface, whereas on the remaining adsorption sites water adsorbes molecularly. In a previous work by our group,[12] we have investigate the behaviour of water multilayers with increasing thickness up to 12 nm comparing density functional tight binding results with molecular mechanics molecular dynamics simulations.

However, the classical model that we used,[12] although catching the general aspects of the water structure and of solvation, has shown limited accuracy in the description of the details of the water coordination to the exposed surface undercoordinated iron sites. In the abovementioned model, we observed longer distances (~2.7-2.8Å) between the superficial iron atoms and the oxygen atoms of adsorbed water molecules (from now on Fe-$O_{water}$) compared to higher-level calculations using hybrid DFT (short for density-functional theory) and Hubbard corrected SCC-DFTB (short for self-consistent charge density-functional tight-binding), from now on DFTB+U, methods that predict Fe-$O_{water}$ distances about 2.2 Å.

Herein, we provide an optimized set of CLASS2 force field parameters that corrects this issue and, therefore, accurately describes at the classical level the water coordination on the Fe sites over the partially hydroxylated $Fe_3O_4$(001) surface. In addition, we observe a quantitative agreement of surface and bulk properties between classical molecular-mechanics molecular-dynamics (from now on classical MM-MD) simulations and DFTB+U molecular dynamics (from now on DFTB-MD) simulations for a bulk water density distribution along the $Fe_3O_4$(001) surface normal. Finally, we test the transferability of the optimized parameters for the description of the water adsorption on the curved surface of a spherical $Fe_3O_4$ nanoparticle, from now on nanosphere (NS), of realistic size with a diameter of 2.5 nm.

## 2. Computational Methods

All atomistic MM-MD simulations were carried out with the LAMMPS program (version 7 Aug 2019).[13] We made use of the CLASS2 potentials (see Ref. 14 for a full description of the COMPASS/CLASS2 force field (FF)). This FF describes the non-bonded interactions for the repulsive and dispersive van der Waals forces through a Lennard-Jones 9-6 potential form (eq. 1), whereas the long-range electrostatic interactions are modeled by a classical Coulomb functional form (eq. 2).

$$E_{vdW} = \sum_{i,j} \epsilon_{ij} \left[ 2\left(\frac{\sigma_{ij}^0}{r_{ij}}\right)^9 - 3\left(\frac{\sigma_{ij}^0}{r_{ij}}\right)^6 \right] \quad (1)$$

$$E_{elec} = \frac{1}{4\pi\epsilon_0} \sum_{i,j} \frac{q_i q_j}{r_{ij}} \quad (2)$$

Here, $\sigma_{ij}$ defines the inter-atomic distance between a pair of atoms at which the potential energy function assumes a minimum value, and $\varepsilon_{ij}$ defines the potential well depth for this pairwise potential. $q_i$ and $q_j$ are the partial atomic charges on the atoms i and j. Bonded and non-bonded parameters for the CLASS2-based three-site water model and the hydroxyl group are taken from the INTERFACE-FF.[15] The LJ(9-6) parameters for the Fe(II), Fe(III), and O(II) atom-types were taken from Ref. [16], and the partial-atomic charges for these atoms were derived from the DFT/HSE06 calculations. For unlike atom-types, the sigma and epsilon cross-parameters are given by the 6[th] power combining rules[14] accordingly to Eq. 3 and 4, respectively:

$$\sigma_{i,j} = \left[\frac{(\sigma_i^0)^6 + (\sigma_j^0)^6}{2}\right]^{\frac{1}{6}} f_1 \quad (3)$$

$$\epsilon_{i,j} = 2\sqrt{\epsilon_i \epsilon_j} \left[\frac{(\sigma_i^0 \sigma_j^0)^3}{(\sigma_i^0)^6 + (\sigma_j^0)^6}\right] f_2 \quad (4)$$

The scanning of both epsilon and sigma Fe-O$_w$ cross-parameters consist of systematic variation in the $f_1$ (Eq. 3) and $f_2$ (Eq. 4) factors over their original values (case in which $f_1$ and $f_2$ are equal to the unit) until the error function reach satisfactory agreement against the reference data set. This protocol is in line with a previous published protocol by one of the authors.[17]

$$RMSE = \sqrt{\frac{\sum_{i=1}^{n}(X_{MM,i} - X_{QM,i})^2}{n}} \quad (5)$$

Values close to zero indicate good agreement, whereas deviations to larger values indicate a more flawed agreement between the MM ($X_{MM,i}$) and the QM-reference ($X_{QM,i}$) predictions for the peaks maximum position in the number density profiles. To accurately determine the QM reference data in Eq. 5, we estimate, through a non-linear Gaussian fitting, the first, second, and third maxima peaks in the DFTB-MD density profiles. The atomic coordinates of the Fe, O and H in the $Fe_3O_4$ slab and NS are freeze at the DFTB+U-optimized geometry by zeroing the forces on these atoms every MM-MD simulation step. To avoid spurius effects due to water evaporation at the trilayer/vacuum interphase in the MM-MD simulations, we included sufficient water molecules between the third water layer and the vacuum phase to keep the inner solvation layers stable on the surface. To solvate this system, we made use of the PACKMOL program[18] to randomly displace the water molecules on both sides of the hydroxylated $Fe_3O_4$ slab by a ~12 nm-thick water multilayer, whose images are periodically repeated along the z-direction. To get a better molecular packing than what provided by PACKMOL, we carried out an energy minimization calculation, where we allowed to relax the water molecules sandwiched between the opposite sides of the $Fe_3O_4$ slab (freezed at the original DFTB atomic positions of $Fe_3O_4$ atoms) and the cell dimension along the z-direction. We applied a stress tensor at P = 1 atm for the virial component of the pressure (the non-kinetic portion) along the z-component of the simulation box with a threshold for forces of $1\times10^{-5}$. On the xy-plane parallel to the $Fe_3O_4$ slab, a standard energy minimization of the water molecules, through the conjugate gradient algorithm with a threshold for forces of $1\times10^{-5}$, was performed. After this relaxation, the bulk-water dimension along the z-direction slightly shrinks of ~0.9%. The final bulk-water density is 0.997 g/cm$^3$, in excellent agreement with the experimental estimation of water density at T = 300 K and P = 1 atm (0.996 g/cm³). The equilibration phase was carried out for 10 ns in the isotherm-isobaric (NVT) ensemble until convergence of the bulk-water density at T=300 K. The production phase explored 50 ns of the phase space in the NVT ensemble at T=300 K. The long-range solver particle-particle particle-mesh[19] handled the electrostatic interactions with a real space cutoff of 10 Å and a threshold of $10^{-6}$ for the error tolerance in forces. For the short-range LJ (9-6) potential, we used a cutoff of 10 Å. The Newton's equations of motion were solved using the Velocity-Verlet integrator with a time step of 1.0 fs.

### *2.1 Potential energy function (PEF) calculations*

Density functional theory calculations of a single water molecule adsorbed on bare $Fe_3O_4$(001) surface were used as a first benchmark to tune the force field parameters. The system was allowed to ionically relax to a stable configuration. The adsorption of water oxygen ($O_{water}$) on top of a surface 5-coordinated Fe ion at octahedral site was observed in agreement with a previous study.[11] Therefore, DFT/HSE06 calculations were performed for 23 configurations of the water molecule at different distances from the surface Fe. For this purpose, the adsorbed water molecule has been rigidly shifted along the surface normal. The adsorption energies of these configurations were obtained performing Self-Consistent Field (SCF) calculations without any ionic relaxation. Then, a distance-dependent (Fe–$O_{water}$) potential energy function (PEF) was built up.

*2.2 Density profile calculations*

For the second refinement of FF parameters, linear number density profiles (atoms/Å) were calculated to fit the MM-MD and DFTB-MD results, in which only O atoms belonging to the molecular water were considered. First, we have divided the space along the z coordinate in equally sized bins (Δz) of thickness set at 0.1 Å. Then, the particles count for each bin was normalized by the total count of particles and by its size.

*2.3 DFT and DFTB computational details*

To tune the force field parameters, hybrid density functional theory calculations (HSE06) were carried out using the CRYSTAL17 package.[20,21] In these calculations, the Kohn–Sham orbitals are expanded in Gaussian-type orbitals (the all-electron basis sets are H|511G(p1), O|8411G(d1) and Fe|86411G(d41), according to the scheme previously used for $Fe_3O_4$).[9,22] The convergence criterion of 0.023 eV/Å for forces was used during geometry optimization and the convergence criterion for total energy was set at $10^{-6}$ Hartree for all the calculations. The k points generated by the Monkhorst–Pack scheme were chosen to be 3×3×1 since total energy difference was found to be below 1 meV when compared with larger grids up to 6×6×1. According to a previous report,[23] the inclusion of the van der Waals correction (DFT+D2)[24] only slightly changes the adsorption energy of water on the $Fe_3O_4$(110) surface, so no van der Waals correction is included in this work.

To further refine the force field parameters, molecular dynamics simulations are performed using SCC-DFTB method implemented in the DFTB+ package.[25] The SCC-DFTB is an approximated DFT-based method that derives from the second-order expansion of the Kohn-Sham total energy in DFT with respect to the electron density fluctuations. The SCC-DFTB total energy can be defined as:

$$E_{tot} = \sum_{i}^{occ} \varepsilon_i + \frac{1}{2}\sum_{\alpha,\beta}^{N} \gamma_{\alpha\beta}\Delta q_\alpha \Delta q_\beta + E_{rep} \quad (6)$$

where the first term is the sum of the one-electron energies $\varepsilon_i$ coming from the diagonalization of an approximated Hamiltonian matrix, $\Delta q_\alpha$ and $\Delta q_\beta$ are the induced charges on the atoms α and β, respectively, and $\gamma_{\alpha\beta}$ is a Coulombic-like interaction potential. $E_{rep}$ is a short-range pairwise repulsive potential. More details about the SCC-DFTB method can be found in Refs.[26,27,28]. DFTB will be used as a shorthand for SCC-DFTB.

For the Fe-Fe and Fe-H interactions, we used the "trans3d-0-1" set of parameters, as reported previously.[29] For the O-O, H-O and H-H interactions we used the "mio-1-1" set of parameters.[30] For the Fe-O interactions, we used the Slater-Koster files fitted by us previously,[31] which can well reproduce the results for magnetite bulk and surfaces from HSE06 and PBE+U calculations. To properly deal with the strong correlation effects among Fe 3d electrons,[32] DFTB+U with an effective U-J value of 3.5 eV was adopted according to our previous work on magnetite.[9,12,22,31,33] The convergence criterion of $10^{-4}$ a.u. for forces was used during geometry optimization together with conjugate gradient optimization algorithm. The convergence threshold on the self-consistent charge (SCC) procedure was set to be $10^{-5}$ a.u. for flat $Fe_3O_4$(001) surface calculations and $5\times10^{-3}$ a.u. for $Fe_3O_4$ NS calculations. The k points generated by the Monkhorst–Pack scheme were chosen to be 6×6×1 for flat $Fe_3O_4$(001) surface calculations. We further checked that the structure for the adsorption of water molecules is not affected by the inclusion of the van der Waals correction (DFTB+D3).[34,35] Since the variations are within 0.1 Å, no correction will be presented in the following.

DFTB+U molecular dynamics were performed within the canonical ensemble (NVT) using an Andersen thermostat to keep the temperature constant at 300 K. The total simulation time is 50 ps with a time step of 1 fs. The convergence threshold on the self-consistent charge (SCC) procedure was set to be $5\times10^{-3}$ a.u. and the k points generated by the Monkhorst–Pack scheme were chosen to be 4×4×1 (since total energy difference was found to be below 1 meV when compared with larger grids up to 6×6×1). To well describe the hydrogen bonds, a modified hydrogen bonding damping (HBD) function was introduced with a $\zeta$ = 4 parameter.[36]

For the $Fe_3O_4$(001) surface, we use the SCV model. According to previous reports, this structural model is more stable than other models.[37] We used the same structure presented in our previous works,[9,12,31] which is a (1×1) 17-layer slab with inversion symmetry. In the z-direction a vacuum of more than 12 Å was introduced to avoid the spurious interaction between the periodic sides of the slabs. Five layers in the middle of the slab are kept fixed to the bulk positions, whereas the other layers are fully relaxed. For water adsorption, molecules were put on both sides of the slab. For the Fe$_3$O$_4$ NS, we use the model (with about 1000 atoms) proposed in our previous work.[33]

To evaluate the stability of one water molecule adsorbed on the Fe$_3$O$_4$(001) surface and on the Fe$_3$O$_4$ NS, the adsorption energy ($E_{ads}$) was calculated as follows:

$$E_{ads} = (E_{total} - E_{slab/NS} - E_{H_2O}) \quad (7)$$

where $E_{total}$ is the total energy of the whole system (surface slab/NS and adsorbed water), $E_{slab/NS}$ is the energy of the Fe$_3$O$_4$(001) surface slab or the energy of the Fe$_3$O$_4$ NS and $E_{H_2O}$ is the energy of one isolated water molecule.

### 3. Results and Discussion

*3.1 Fitting of Potential Energy Function (PEF) from hybrid DFT(HSE06) calculations for one water molecule adsorbed on Fe$_3$O$_4$(001) surface*

As stated earlier, the original sets of CLASS2-FF parameters, adopted in Ref. 11, by merely applying the sixth-power combining rules to describe the cross-interaction between this pair of unlike atoms (see Eq. 3 and Eq. 4), led to an overestimated Fe-O$_{water}$ distance. To tackle this issue, we carry out a systematic *ad hoc* parametrization of both the repulsive (step 1) and attractive (step 2) components in the Lennard Jones (9-6) (from now on LJ, see Eq. 1 in Section 2) potential for this pair of unlike atoms. This procedure consists of an iterative two-step optimization (see the Supplementary Material for more details regarding the parametrization protocol): 1) Estimation of the potential energy function (from now on PEF) for the adsorption of a single water molecule on the bare Fe$_3$O$_4$(001) surface, seeking for the best agreement between the hybrid DFT (HSE06) and the classical results. 2) Fine-tuning of the epsilon cross-parameters (ε[Fe-O$_{water}$]) in a systematic fashion to find the best match between classical MM-MD and DFTB-MD predictions for the water-trilayer density profile. We construct an objective error function (Eq. 5 in Section 2) to measure the agreement between the QM (quantum mechanics) reference data set and the corresponding classical MM-MD results.

Steps 1 and 2 are repeated until a satisfactory agreement between the classical MM-MD and the DFTB-MD predictions. Since the accuracy of the latter method has been already validated against hybrid DFT calculations in our previous publication,[11] we use DFTB-MD results as the reference data set for the water density profile. Moreover, such a large number of atoms is unpractical for standard higher-level calculations (e.g., ab initio DFT-MD simulations). It is also important to mention that partial atomic charges on the classical atoms are derived based on the DFT level of theory and are kept fixed during all this optimization procedure (see Table S3 in the Supplementary Material).

Figure 1 shows the PEF for the adsorption of a single water molecule on the bare $Fe_3O_4$(001) surface. Starting from the DFT(HSE06)-optimized structure, the water molecule is rigidly shifted along the normal to the surface. To fit the target QM data, we carry out systematic scanning by employing multiplicative factors over the original set of CLASS2-FF parameters (brown curve), taken by us as starting-point in this optimization procedure. More details can be found in the Supplementary Material.

We first fine-tune the sigma cross-parameter (with the epsilon parameter given by Eq. 4) to determine the best fit for the inter-atomic distance between the Fe-$O_{water}$ pair of atoms. Once a good match for the position of the PEFs minimum by QM and by MM is observed (Figure 1), we fine-tune only the epsilon cross-parameter, to adjust the well depth to the classical LJ potential.

*3.2 Comparing MM-MD density profile with DFTB-MD calculations for a water trilayer on $Fe_3O_4$(001) surface*

To assess the performance of this prime set of CLASS2-FF parameters obtained from the fitting of the DFT adsorption PEF (step 1), we carried out classical MM-MD simulations for the trilayer model system. We still observe unsatisfactory agreement between the classical MM-MD predictions and the DFTB-MD target data for the water-trilayer density profile along the $Fe_3O_4$(001) surface normal (Figure S1, Item B). Although it shows a definite improvement over that obtained by the original CLASS2-FF parameters in our previous publication,[11] we decide to carry out a further step of refinement to enhance the agreement between the classical MM-MD results and the DFTB-MD reference data set. In this second-step parametrization, we carry out a fine-tuning of the epsilon cross-parameter between the Fe and $O_{water}$ atoms.

Figure 2 shows the water density profiles along the $Fe_3O_4$(001) surface normal. Classical MM-MD predictions using the final parametrized set of CLASS2-FF parameters are shown in cyan (trilayer model) and green (bulk-water model). The reference data for the water density profile predicted by the DFTB-MD simulations is shown in black. For the sake of comparison, we also include the water density profile estimated by the original model (blue) for the trilayer model in our previous publication.[12] The first two peaks for the MM-MD simulation with the original CLASS2-FF parameters merge into one single peak just above 10 Å for the simulation with the optimized parameters, in agreement with DFTB-MD results, thanks to the improved description of the Fe-O$_{water}$ cross-interaction.

### 3.3 MM-MD density profile for bulk water on Fe$_3$O$_4$(001) surface

To further assess the performance of this final set of CLASS2-FF parameters (see Table S3 in the Supplementary Material), we built up a Fe$_3$O$_4$(001)/water interface model of realistic scale and density (bulk-water phase above and below the partially hydroxylated Fe$_3$O$_4$(001) slab of 80×80Å, used for the trilayer model). We then check whether the presence of bulk-water could somehow affect the density of water molecules along the Fe$_3$O$_4$ (001) surface normal compared with the trilayer model. There are no substantial differences between the Fe$_3$O$_4$/trilayer-water (Figure 2, cyan) and the Fe$_3$O$_4$/bulk-water (Figure 2, green) MM-MD predictions for the water density profile, although a slight shift towards the surface is observed in the latter model.

Furthermore, we simulate an even larger model 160×160Å of a partially hydroxylated Fe$_3$O$_4$ surface containing a total of 112624 water molecules. To the best of our knowledge, this system size is beyond any previous classical MM-MD simulation for the Fe$_3$O$_4$(001) surface/water interface to date. We notice that the most prominent finite-size effect is an induced ordering, mainly in the first solvation layer, as a consequence of the enlargement of the Fe$_3$O$_4$ surface area. This is evident in Figure S2 in the Supplementary Material, where we observe a higher peak intensity for the Fe$_3$O$_4$ 160×160 Å surface (orange profile). Moreover, we observe no substantial difference of the farther solvation layers in the bulk-water above the Fe$_3$O$_4$ surfaces.

### 3.4 Assessment of optimized FF parameters on single water molecule adsorption on Fe$_3$O$_4$(001) surface

The results above indicate that a Fe$_3$O$_4$ surface/water interface model, having both a realistic system size and bulk-water density, could take advantage of these new CLASS2-FF parameters. To corroborate this last assumption, we finally compare structural parameters of the optimized geometry of one water molecule adsorbed on bare and partially hydroxylated Fe$_3$O$_4$(001) surfaces at different levels of theory (see Figures 3(a) and 3(b), respectively). Table 1 shows that the interatomic distances from the optimized set of MM parameters are in better agreement with DFTB and DFT results compared to the original set. In particular, for the bare surface, the Fe-O$_{water}$ distance was 2.84 Å with the original FF, which is quite too long with respect the DFT(HSE06) reference value of 2.16 Å, as we pointed out in our previous work,[11] whereas the same distance becomes 2.01 Å with optimized FF set of parameters, which is only 6.9% shorter than the reference value. Table S2 in the Supplementary Material shows that the adsorption energy obtained from the optimized set of FF parameters is in better agreement with DFTB and DFT data compared to the original FF. For instance, we observe adsorption energy of -0.66 eV using the original set of FF parameters, which is off by 43% compared to the DFT(HSE06) reference data of -1.15 eV, while the adsorption energy predicted by the optimized FF matches exactly the DFT(HSE06) reference value.

*3.5 Assessment of the force fields for their use with spherical $Fe_3O_4$ nanoparticles.*

In this section, we present the assessment of the transferability of the newly optimized set of parameters for the description of water adsorption on a $Fe_3O_4$ NS of realistic size with about 1000 atoms (see Figure 4).

To this end, we compare the adsorption structure and energy of a single water molecule on different sites of the partially hydroxylated $Fe_3O_4$ NS, as obtained at the MM level of theory with the CLASS2 force field parameters with the results from a quantum mechanical QM method. We use DFTB+U results as the QM reference data set, since DFTB+U was previously validated against higher-level DFT results.[12] The MM results are obtained both through the direct optimization of DFTB+U structures (second rows of Tables 2 and 3) and through the final optimization after 50 ns of MM-MD simulation at 300 K (third rows of Tables 2 and 3). In all the MM calculations, the structure of the partially hydroxylated $Fe_3O_4$ NS is kept fixed at DFTB+U optimized geometry before water adsorption.

When one water molecule is adsorbed on sites 1, 2 and 3 (see Figure 5), MM structural results, both when optimized before and after molecular dynamics, very much resemble the DFTB ones. For the adsorption on sites 4 and 5, classical MM results before the MM-MD are in good agreement to DFTB ones (see Figures 6(a) and 6(c)), while classical MM results after MD are slightly different (see Figure 6(b) and 6(d)). Figures 6(a) and 6(c) show that before the MM-MD simulations, the water molecule tends to form one hydrogen bond, whereas Figures 6(b) and 6(d) show that after the MM-MD simulations, the water molecule interacts with two O atoms on the NS forming two H-bonds.

Table 3 reports the water binding energy values. We observe a good agreement of the binding energies at DFTB+U and MM levels of theory. We also observe that the optimized configuration after the MM-MD simulation (third row) is more stable than the one optimized before the MM-MD (second row). We may notice that, due to the formation of a second H-bond after MM-MD, the water molecule undergoes a structural rearrangement. The Fe-$O_{water}$ distance is not much affected, whereas the molecule rotates to allow the formation of the new H-bonds, resulting in a significantly reduced distance between the $O_{water}$ atom and the NS centroid (see second and third rows in Tables 2).

According to this assessment analysis, the transferability of the optimized CLASS2 set of FF parameters turns out to be less accurate as the Fe neighborhood becomes less similar to that of the $Fe_3O_4$(001) flat surface for which the parameters have been optimized. The MM-MD simulations, depending on the site morphology, tend to maximize the number of H-bonds established by the water molecule with the nanoparticle surface atoms, as discussed above. We do not observe this phenomenon in the classical $Fe_3O_4$(001) flat surface model because of the lower availability of neighboring surface O atoms as H-bond acceptors, as evident by Figure 7(a). In particular, Figure 7(a) shows that the adsorption site on the $Fe_3O_4$(001) flat surface is morphologically similar to the adsorption sites 1, 2 and 3 of the spherical nanoparticle in Figure 5. On the contrary, the morphology of site 5 in Figure 6(c) and 6(d), as in the case of site 4 in Figure 6(a) and 6(b), is different to that of the flat surface. Here, several surface O atoms are available as H-bond acceptors and the water molecule can thus establish strong H-bonds (with an O-H--O angle of ~170° close to the optimal one of 180°).

We wish to add that for all the DFTB+U calculations, we made use of the HBD correction to get the best possible description of the H-bond interactions.[36] Based on the results above, we conclude that the classical model tends to overestimate the H-bond interaction between the water molecule and the nanoparticle surface.

To conclude this section, based on the results presented and discussed above, we consider that the optimized set of CLASS2-FF parameters obtained in Section 3.1 are suitable for the description of the water interface with a spherical $Fe_3O_4$ nanoparticle, because they perform very well, with respect to the QM reference, in the description of the water adsorption on the surface of spherical nanoparticle, especially when the adsorption site is morphologically similar to that of the flat (001) surface. They are still satisfactorily accurate, even when the adsorption site is more affected by the curvature, except for a rigid shift of the water O atom towards the center of the nanoparticle due to a molecular rotation to achieve the highest number of H-bonds with the surface, keeping the Fe-$O_{water}$ distance essentially unmodified.

## 4 Conclusions

In the present work, we make available a new set of CLASS2-FF parameters (see the Supplementary Material) that accurately describe the molecular interaction between a partially hydroxylated $Fe_3O_4$ (001) surface and the interfacial/bulk-water molecules. Further development towards a combination between both the present CLASS2-FF parameters and the well-established CLASS2-FF parameters for $Fe_3O_4$/organic interfaces[12], open up many possibilities on modeling and simulation of more complex $Fe_3O_4$/organic/water interfaces at realistic time and length scales. Here we have proved their satisfactory transferability to the description of water adsorption on the curved surface of a spherical $Fe_3O_4$ nanoparticle of realistic size (2.5 nm).

**Supplementary Material**

See the supplementary material for further details regarding the parametrization protocol, the optimized LJ-parameters for the Fe–O$_{water}$ cross-interaction together with partial atomic charges for magnetite atoms, and the linear number density profile for the (160 × 160) Å$^2$ Fe3O4/bulk-water system.

**Data availability statement**

The data that support the findings of this study are available from the corresponding author upon reasonable request.

**Acknowledgments**

The authors are grateful to Federico Soria for the useful discussions and to Lorenzo Ferraro for his technical help. The project has received funding from the European Research Council (ERC) under the European Union's HORIZON2020 research and innovation programme (ERC Grant Agreement No [647020]).

This paper is dedicated, on the occasion of her 70$^{th}$ birthday, to Prof. Annabella Selloni, who performed groundbreaking work on the theoretical simulation of semiconducting oxides.

**Table 1.** Structural parameters of one water molecule adsorbed on bare and partially hydroxylated $Fe_3O_4$ (001) surfaces. The inter-atomic distance obtained by DFT/HSE06 method is taken as the reference for calculating the error.

| Inter-atomic distance (Å) | DFTB+U | DFT/HSE06 | MM (original) | MM (optimized) |
|---|---|---|---|---|
| Fe - $OH_2$[a] | 2.23 | 2.16 | 2.84 (31.5%) | 2.01 (-6.9%) |
| Fe - OH[b] | 1.89 | 1.93 | 1.89[c] | 1.89[c] |
| Fe - $OH_2$[b] | 2.01 | 2.06 | 2.69 (30.6%) | 1.95 (-5.3%) |
| HO- -HOH[b] | 1.57 | 1.48 | 1.62 (9.5%) | 1.54 (4.1%) |
| HO- -$OH_2$[b] | 2.57 | 2.52 | 2.60 (3.2%) | 2.53 (0.4%) |

[a] Bare surface (see Figure 3(a))
[b] Partially hydroxylated surface (see Figure 3(b))
[c] Fixed atoms in the slab.

**Table 2.** Fe-$O_{water}$ inter-atomic distance (Å) of one water molecule adsorbed on different sites of a partially hydroxylated $Fe_3O_4$ NS at DFTB+U and MM levels of theory. The distance between the water oxygen atom and the center of the NS is reported in parenthesis.

| Fe-$OH_2$ (Å) | Site 1 | Site 2 | Site 3 | Site 4 | Site 5 |
|---|---|---|---|---|---|
| DFTB+U | 2.09 (15.06) | 1.94 (15.05) | 2.02 (13.83) | 1.99 (13.41) | 2.01 (13.29) |
| MM (from DFTB+U) | 2.08 (15.02) | 1.94 (15.01) | 2.04 (13.82) | 1.99 (13.41) | 1.94 (13.31) |
| MM (from MM-MD) | - | - | - | 2.10 (13.05) | 2.07 (13.01) |

**Table 3.** Binding energies (eV) of one water molecule adsorbed on different sites of a partially hydroxylated $Fe_3O_4$ NS at DFTB+U and MM levels of theory.

| $E_{binding}$ (eV) | Site 1 | Site 2 | Site 3 | Site 4 | Site 5 |
|---|---|---|---|---|---|
| DFTB+U | -1.03 | -2.71 | -1.35 | -1.93 | -1.54 |
| MM (from DFTB+U) | -1.14 | -1.97 | -0.80 | -2.00 | -1.92 |
| MM (from MM-MD) | - | - | - | -2.59 | -2.78 |

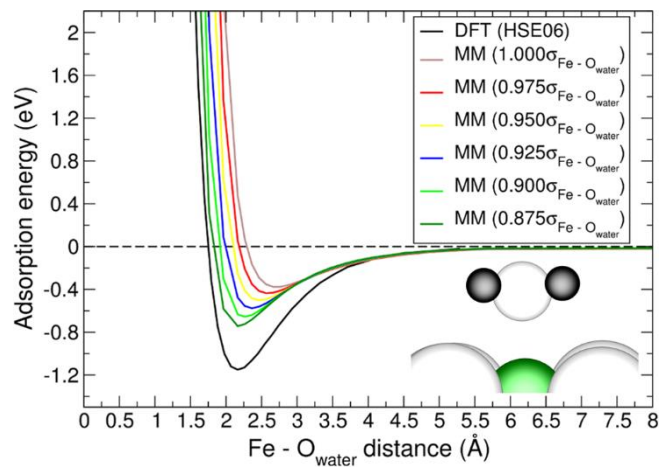

**Figure 1.** PEFs for the adsorption of a single water molecule on the bare $Fe_3O_4$ (001) surface calculated at the DFT(HSE06) and the classical level of theory. The DFT curve is shown in black. All other colors represent each set of CLASS2 parameters and their respective profiles obtained by scanning up the cross-parameters epsilon and sigma between the Fe-$O_{water}$ atoms. The inset shows the side view of the single water molecule adsorbed on the surface. The green, black, small and large white beads represent Fe, H, O from water and O from the surface, respectively.

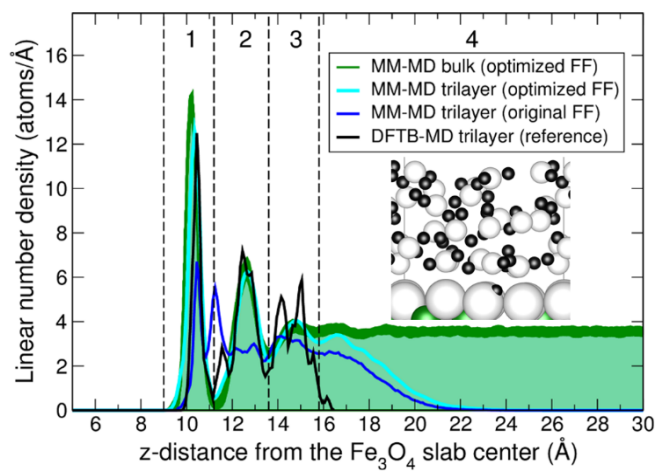

**Figure 2.** Number density profiles of water molecules along the partially hydroxylated $Fe_3O_4$ (001) surface normal. DFTB-MD simulation (black), classical MM-MD bulk-water simulation (green, thick line), classical MM-MD trilayer simulation with the optimized CLASS2-FF (cyan), and classical MM-MD trilayer simulation with the original CLASS2-FF (blue). Numbers between the dashed lines stand for different hydration shells. The inset shows the DFTB-MD water-trilayer model. The green, black, small and large white beads represent Fe, H, O from water and O from the surface, respectively.

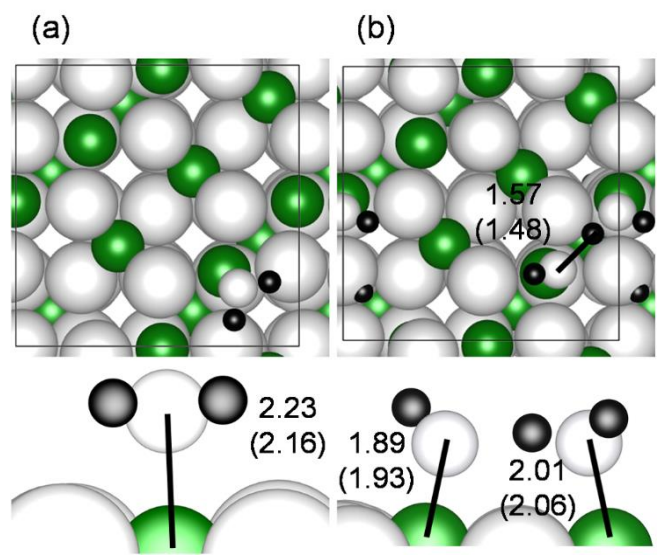

**Figure 3.** Top and side view of a single water molecule adsorbed on (a) bare $Fe_3O_4$(001) surface and on (b) partially hydroxylated $Fe_3O_4$(001) surface. The green, black, small and large white beads represent Fe, H, O from water and O from the surface, respectively. The hydrogen and Fe-$O_{water}$ bonds are indicated by the solid lines. The bond length calculated by DFTB+U (outside the round brackets) and DFT/HSE06 (inside the round brackets) are given for each configuration. The black squares represent the ($\sqrt{2} \times \sqrt{2}$)R45° unit cell used in the calculations.

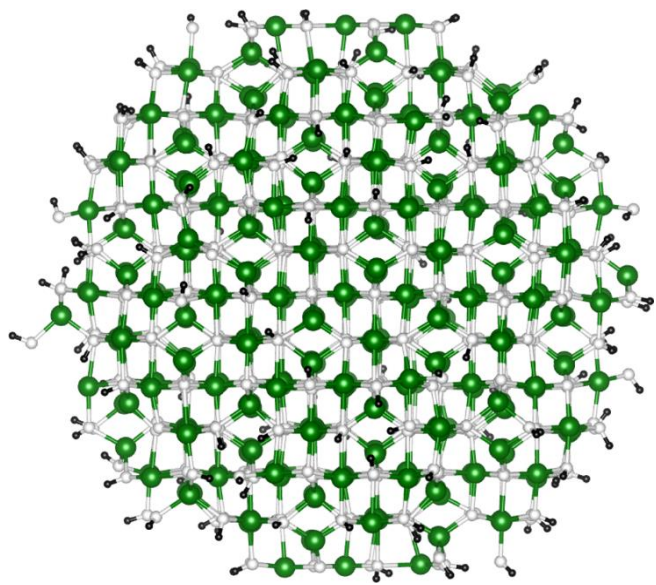

**Figure 4.** Partially hydroxylated spherical $Fe_3O_4$ nanoparticle model of about 1000 atoms with a diameter of 2.5 nm. The green, black and white beads represent Fe, H and O, respectively.

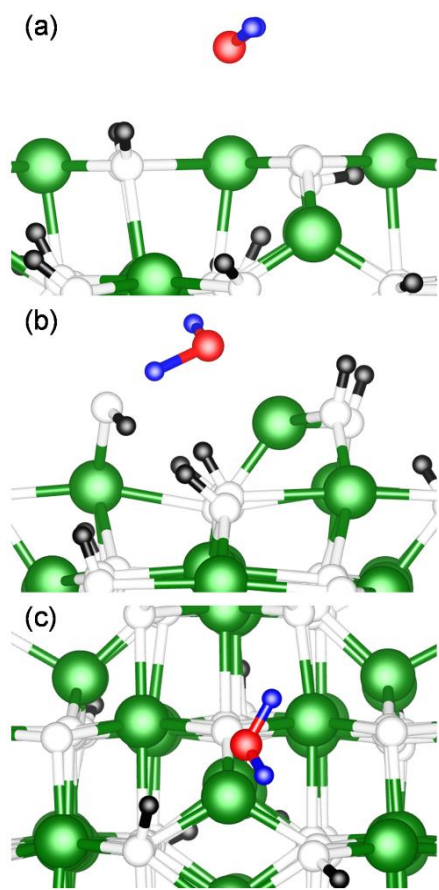

**Figure 5.** Adsorption of one water molecule on (a) site 1, (b) site 2 and (c) site 3 at DFTB+U level. Structures optimized at MM level starting from DFTB+U ones are found to be nearly identical and are not reported. The green, black, white, blue and red beads represent Fe, H, O from the partially hydroxylated NS and H, O from the water molecule, respectively.

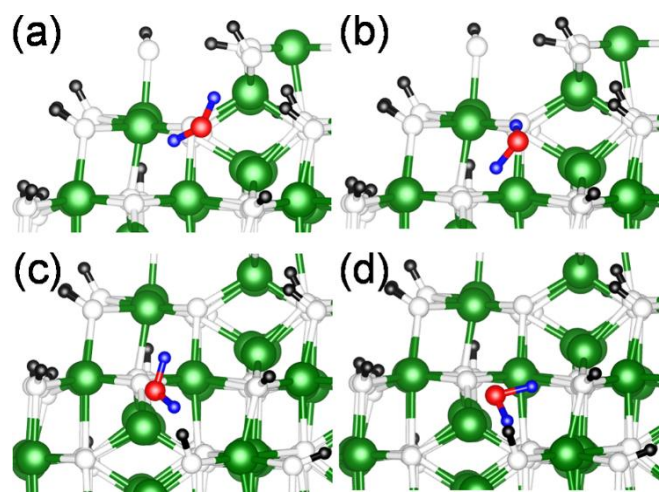

**Figure 6.** Adsorption of one water molecule on site 4 (a, b) and on site 5 (c, d) at DFTB+U level of theory on the left and at MM (optimized geometry after MM-MD) level of theory on the right. Structures optimized at MM level starting from DFTB+U ones are nearly identical and not reported. The green, black, white, blue and red beads represent Fe, H, O from the partially hydroxylated NS and H, O from the water molecule, respectively.

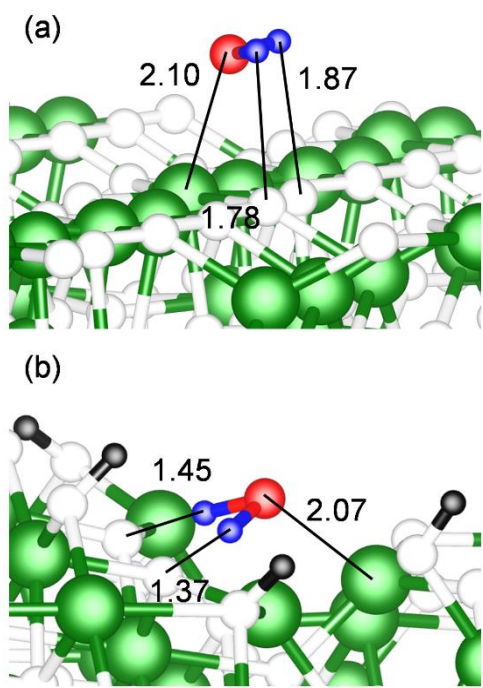

**Figure 7.** Adsorption of one water molecule on (a) the $Fe_3O_4$ (001) surface and (b) the site 5 of the partially hydroxylated $Fe_3O_4$ NS at the MM level of theory. Bonds are indicated by the solid black lines. The green, black, white, blue and red beads represent Fe, H, O from the partially hydroxylated NS and H, O from the water molecule, respectively.